# INFORMATION RETRIEVAL FROM INTERNET APPLICATIONS FOR DIGITAL FORENSIC

Ipsita Mohanty and R. Leela Velusamy

Department of Computer Science and Engineering, National Institute of Technology, Tiruchirappalli, India
ipsita.mohanty689@gmail.com
leela@nitt.edu

## ABSTRACT

*Advanced internet technologies providing services like e-mail, social networking, online banking, online shopping etc., have made day-to-day activities simple and convenient. Increasing dependency on the internet, convenience, and decreasing cost of electronic devices have resulted in frequent use of online services. However, increased indulgence over the internet has also accelerated the pace of digital crimes. The increase in number and complexity of digital crimes has caught the attention of forensic investigators. The Digital Investigators are faced with the challenge of gathering accurate digital evidence from as many sources as possible. In this paper, an attempt was made to recover digital evidence from a system's RAM in the form of information about the most recent browsing session of the user. Four different applications were chosen and the experiment was conducted across two browsers. It was found that crucial information about the target user such as, user name, passwords, etc., was recoverable.*

## KEYWORDS

*Digital forensic, Digital evidence, Live acquisition, Internet application*

## 1. INTRODUCTION

Digital forensics is a branch of forensic science encompassing the recovery and investigation of material found in digital devices, often in relation to computer crime [1]. It involves application of scientific methods within the regulations of law [2] [3]. At the most basic level, digital forensic is the process of acquiring, analyzing, and presenting the digital evidence [4]. Digital evidence is the information collected from digital media involved in crime, such as CDs, DVDs, flash drives, floppy disks, memory cards, mobile phones, network devices, RAM, etc., [2]. It is the basis upon which an assertion is established. Acquisition and Analysis of digital evidence has become an intensive area of research due to the increasing frequency of digital crimes across the world.

For crime investigation, the data stored in target user's system is of great significance. These data can be either static or live. Static data is stored in static storage devices such as hard disk, CDs, flash drives, etc., whereas live data is stored in RAM [2]. Live data, unlike static, changes continuously but contains the current information about the system. Any application used in a system gets loaded into RAM for operation. So, the content of RAM holds the key to information about the applications used by the user on the target system. Valuable information which can be obtained from the RAM includes the processes running, ports opened, files opened for each process, user names and passwords of the user's accounts (created for different online applications and system log on), chat contents, e-mails, contacts, etc. Since the user names and passwords are recoverable, the investigator can log in to the respective accounts and collect more detailed information. A hit-and-trial method may be further adopted across multiple applications to check whether the same user name and password allows access or not.

This enables the investigator to collect information from other online sources which had not been accessed on the target system. Thus, RAM is an important source for collection of live evidence in digital investigation and cannot be ignored.

Simon and Slay were able to retrieve live data such as communication content, communication history, contacts, passwords, and encryption keys for the application Skype [5]. In this paper, the work by Simon and Slay is extended for more diverse internet applications such as social networking, net banking, and online train reservation systems. Besides widening the scope of applications, this paper also takes into consideration multiple browsers to provide a comparative study about how the choice of a browser impacts the amount of data retrieved. The objective of this work is to collect relevant information about the target user from a number of websites that may have been accessed in a particular browsing session. With increasing focus to unveil digital crimes, the approach discussed in this paper acts as a potential tool for gathering live evidence from the target user.

The organization of the remaining portion of the paper is as follows. Section II describes briefly the basic concepts about Digital Forensic. In Section III, the adopted methodologies are described. In Section IV, the results of the analysis are discussed. Finally, conclusion and future work is discussed in section V.

## 2. BACKGROUND

The process involved in Digital Forensic is split into three main phases namely Acquisition, Analysis and Report. Acquisition (imaging) is the process of creating the forensic duplicate i.e., a bit by bit copy of the digital media under investigation [6] [2] [7]. The goal of this phase is to save digital information from all sources possible [8]. However, a step which logically precedes acquisition is identification of various sources of data. Analysis, the second phase, can be defined as the in-depth systematic search of evidence [1]. The third phase, Report, involves complete description of all the actions taken in the first two phases and the conclusion drawn from analysis, so that a proper documentation of the investigation process can be submitted to the court of law. Digital Evidence, being a collection of bits, is very sensitive and can be easily altered [9] [2]. Any scientific procedure adopted during investigation should make no changes to the evidence in order to ensure its admissibility in the court. In case of any alteration due to forensic procedures, a proper explanation must be provided [4].

Acquisition can be done in two ways: Static and Live. Static acquisition involves halting the target system and making a forensically valid copy, or image, of all attached storage media whereas live acquisition involves gathering data while the system is in operation. Static acquisition has certain demerits, such as the need to shut down the system, incomplete evidence and inability to access the static media if encrypted or locked. Live acquisition makes it possible to get a running picture of the system involving information about opened applications, files, ports, running processes, user names, passwords, encryption keys, etc.; where static acquisition fails. However, live acquisition has limitations such as need for administrator level of access, incorrect information from a compromised system, prior installation of hardware to be used such as Tribble and Firewire based devices, overwriting of some useful contents of RAM due to the software's own signature, inconsistent snapshots, and non-repeatable operations. The system state becomes a function of both user and investigator activities. In spite of such shortfalls, live acquisition cannot be avoided since it provides a plethora of information, which static acquisition cannot. Investigators should use softwares which cause as much less modification as possible because acquisition can be done only once though analysis of evidence is repeatable [6] [10]. Modifications can be accepted in critical situations as long as the investigator can clearly validate. Live acquisition is useful when the computer is on (or in standby or sleep mode or locked) and connected to a network [11]. As mentioned by Halderman et al. [12], in these situations RAM contents can be retrieved. But when the system is shut down, only static acquisition can be done. If the system is hibernated, the investigator can get RAM data by imaging the hard disk. Because after hibernation, the contents of RAM get stored in hard disk in a file called "hiberfil.sys". This file can be copied and analyzed to obtain the RAM contents [4]. Imaging of RAM can be done using different tools as discussed by Davis in [13].

The analysis of acquired evidence can be done either through live response or static memory dump analysis. The first approach involves querying the system using API-style tools such as Pslist, ListDlls, Handle, Netstat, Fport, etc. The second approach is to gather useful data from the captured memory image in an isolated manner using different memory analysis tools such as volatility, hex editor and string extraction utilities [14]. Volatility provides command for determining the processes running, the dlls associated with each process, the files opened for each process, the list of opened sockets etc. Hex editor can be used for manual string search. String extraction utilities can be used to extract strings from RAM image which can then be analyzed manually.

Report involves complete documentation of all processes and tools. It also summarizes the conclusions drawn in a layperson's terms [1]. Documentation cannot be considered to be an isolated or specific phase and should be done in every step of the investigation process in order to have a complete description of all steps involved and the results. The prepared document is used for verification and decision making in the court. This also helps a new investigator to understand the whole process quickly with less effort. Since only the investigator can know the evidence in raw level, the way of reporting is very crucial to ensure that others can understand the information from the report easily [9].

This was a brief description about the various steps involved in the digital forensic process. Following these steps, an attempt was made to recover and analyze useful information from browser based applications. Live acquisition was performed by collecting RAM images and analyzing them statically for evidence relevant to the applications used. The obtained information can act as a key to access the target user's profile in multiple sources and collect valuable information about the user's contact, messages exchanged, e-mails etc. In the following Section, the detailed procedure of our work is discussed.

## 3. TESTING PROCEDURE

The internet applications chosen for the testing were: Facebook, Gmail, IRCTC (Indian Railway Catering and Tourism Corporation Limited), and SBI (State Bank of India). The browsers chosen for the experiment were: Mozilla Firefox and Google Chrome. The choice of the applications and the browsers was based on popularity and frequency of usage. Another important factor which helped in selection of applications was the importance of contained data. The aim here was to recover vital information about the target user by leveraging on the RAM content for the most recent browsing session. The testing was carried out individually on each of the applications considered for each of the browsers.

### 3.1. Test Overview

A fresh browsing session (after switching on the computer) was started with no remnant from previous sessions. The application to be tested was opened in the browser and access to internet was obtained by logging into Sonicwall (a firewall interface). The next step was to take images of RAM at different time intervals, trying to cover all critical points without losing any valuable data. During acquisition only the application at issue was opened, in order to avoid alteration of relevant memory contents by other applications. This may not be the case in a real life scenario. The target user might have used more than one application and there is a probability of one application overwriting another application's data. But the testing had to be done in an isolated manner, so as to check for all the probable data that can be retrieved from the application being tested. After acquiring, the images were analysed for contents specific to the application of concern.

### 3.2. Environment Setup

The system being used for testing was a Lenovo 0768 HBQ laptop with following specifications:

- OS: Windows XP Professional, Service Pack 2
- Processor: Intel Pentium Dual-Core Processor T2080 @ 1.73GHz 794MHz
- Physical Memory: 512 MB
- Hard Disk: 80 GB
- Page file size: 0MB
- Internet browser: Mozilla Firefox 6.0 and Google Chrome 19.0.1084.56 m

The page file size was set as zero, in order to have all the contents in RAM, nothing being swapped out to the virtual memory, since the study involved taking image of only RAM. Settings were modified not to save history, where history includes: browsing history, download history, form history, search bar history, cookies, cache content, active logins, offline website data, site preferences, password, and temporary internet files. In Mozilla Firefox, this setting was achieved through private browsing. Permanent private browsing was selected in the privacy panel under options. In Google chrome, this setting was achieved by entering into incognito browsing mode.

### 3.3. Acquisition

For live memory acquisition, the tool 'Nigilant32' [13] [15] was used. This tool need not be installed in the target machine, but can be run from CD or external USB drive. It is just an exe file which needs to be run and has a small footprint, using less than 1 MB in memory, when loaded [13]. It took only 45 seconds to image 512MB of RAM. Although another tool (FTKimager [16]), was also available, Nigilant32 was preferred due to faster response time.

The steps followed in acquisition are:

1. Turn the system on
2. Take image of system memory-Img1.img
3. Start the browser (Mozilla Firefox)
4. Take image of system memory-Img2.img
5. Log in to Sonicwall
6. Take image of system memory-Img3.img
7. Open the application(e.g. Gmail) and log in
8. Take image of system memory-Img4.img
9. Keep the system idle for 1 minute
10. Take image of system memory-Img5.img
11. Keep the system idle for 5 minutes
12. Take image of system memory-Img6.img
13. Log out from the application
14. Take image of system memory-Img7.img
15. Close the browser
16. Take image of system memory-Img8.img
17. Keep the system idle for 1 minute
18. Take image of system memory-Img9.img
19. Log out from Sonicwall
20. Take image of system memory-Img10.img
21. Keep the system idle for 2 minutes

22. Take image of system memory-Img11.img
23. Keep the system idle for 3 minutes
24. Take image of system memory-Img12.img
25. Keep the system idle for 5 minutes
26. Take image of system memory-Img13.img
27. Shut down the system

The above sequence of steps was followed for each browser (Mozilla Firefox and Google Chrome) and all applications under test i.e., Facebook, Gmail, IRCTC, and SBI.

### 3.4. Analysis

In analysis phase, the images taken during acquisition were searched carefully to find information relevant to the application under concern. First, all strings were extracted from the images using Windows Sysinternals utility 'Strings' [17] and stored in different text files for different images. Then the text files were searched to find subtle hints pointing to relevant information like username, password etc. These text files were also used in the plug-in 'strings' of volatility [18] to know about the id of the processes, within which memory space, the strings were stored. The plug-in 'strings' of volatility takes as input an image file and the text file with lines of the form <offset>:<string>, usually created by Sysinternals utility 'Strings' for the same image, and creates a text file containing the corresponding process names (or id of the processes) and virtual addresses for the strings stored in the memory image [19]. The list of running processes while acquiring the image was generated using command 'pslist' of volatility and the pid associated with the searched string was matched to find out the process name.

The images can also be searched for strings using hex editor [20]. But the advantage of using Windows Sysinternals utility 'Strings' is that the output text file contains only printable characters, not the non-printable ones. So it is easy and clear to search.

### 4. RESULTS

The primary data for search were user name and passwords, used for logging in to the applications. The results are summarised in Table 1. This table is followed by the detailed analyses with snapshots for individual applications. The user names and passwords are highlighted in each snapshot.

Table 1 - Presence of Password

| Application ⇨ RAM image ⇩ | Sonicwall | | Facebook | | Gmail | | IRCTC | | SBI(encrypted) | |
|---|---|---|---|---|---|---|---|---|---|---|
| | MF | GC | MF | GC | MF | GC | MF | GC | MF | GC |
| Img1 | No | No | No | No | No | No | No | No | No | No |
| Img2 | No | No | No | No | No | No | No | No | No | No |
| Img3 | Yes | Yes | No | No | No | No | No | No | No | No |
| Img4 | Yes | Yes | Yes | Yes | Yes | Yes | Yes | Yes | No | Yes |
| Img5 | Yes | Yes | Yes | Yes | Yes | Yes | Yes | Yes | No | Yes |
| Img6 | Yes | Yes | Yes | Yes | Yes | Yes | Yes | Yes | No | Yes |
| Img7 | Yes | Yes | Yes | No | Yes | No | Yes | Yes | No | Yes |
| Img8 | Yes | Yes | Yes | No | Yes | No | Yes | Yes | No | No |
| Img9 | Yes | Yes | Yes | No | Yes | No | Yes | Yes | No | No |
| Img10 | No | No | No | No | No | No | No | No | No | No |
| Img11 | No | No | No | No | No | No | No | No | No | No |
| Img12 | No | No | No | No | No | No | No | No | No | No |
| Img13 | No | No | No | No | No | No | No | No | No | No |

**MF: Mozilla Firefox, GC: Google Chrome**

## 4.1. Sonicwall

### 4.1.1. Mozilla Firefox

The user name and password for logging into sonicwall was found in images: Img3 through Img9. The instances were in the memory space of firefox.exe. In Figure 1, a snapshot of the text file img9.txt, created from Img9.img by 'Strings' utility is given, which contains the url of the website, the session id, user name and password of the user. Img3 was acquired after logging into the interface and Img9, just before logging out. After logging out from Sonicwall, the firefox.exe process gets closed. So the contents were not found in the images acquired after that i.e. Img10-Img13.

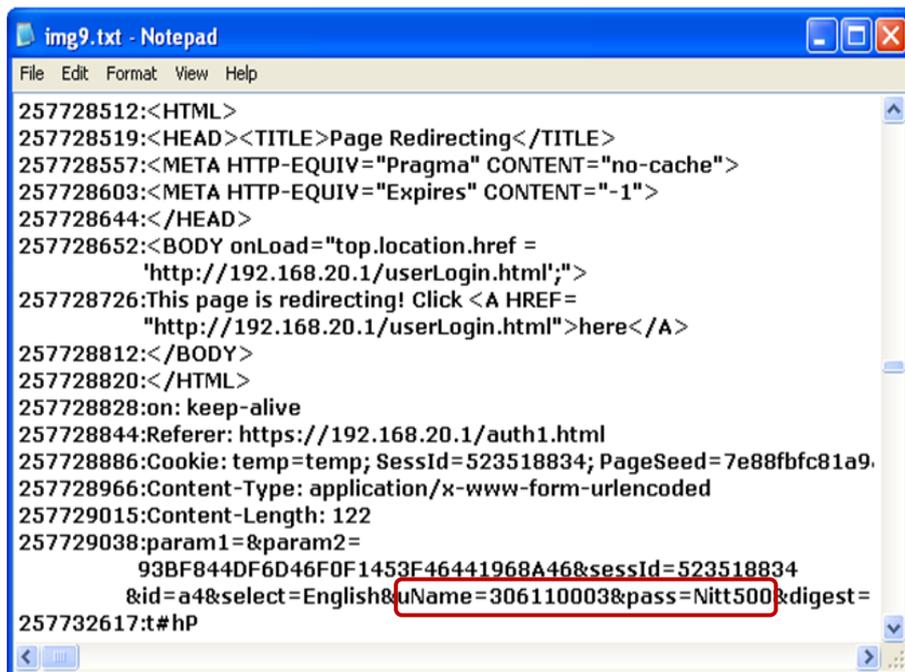

Figure 1. Snapshot of text file from Img9 taken for Sonicwall in Mozilla Firefox

### 4.1.2. Google Chrome

The findings under Google Chrome were similar to those under Mozilla Firefox. The contents of Sonicwall login page: url of the website, session id, user id and password were found in images: Img3 through Img9 i.e. after logging into the interface till logging out from it. The contents were not found in images acquired after log out i.e. Img10-Img13. In Figure 2, a snapshot created from img9.txt is given, which shows the content of the log in page loaded. The instances were in the memory space of chrome.exe and kernel process. The words preceding username and password, which can be used as keywords for search, were 'uName' and 'pass' respectively. These keywords were same for Mozilla Firefox too.

It can therefore be concluded that it is possible to retrieve such important information until the user has not logged out of the interface. This remains constant across the two browsers.

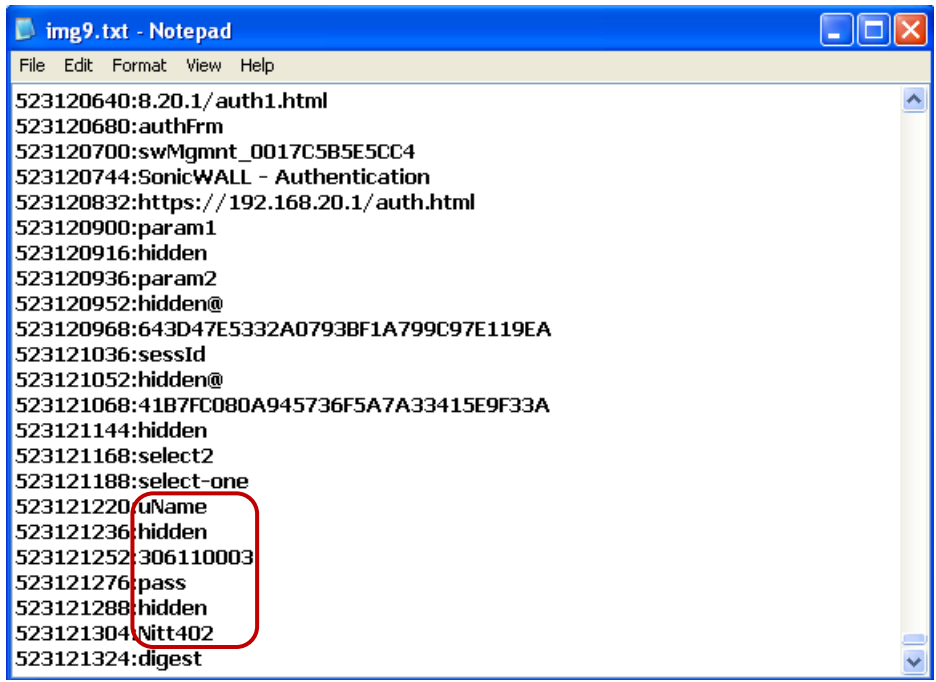

Figure 2. Snapshot of text file from Img9 taken for Sonicwall in Google Chrome

## 4.2. Facebook

### 4.2.1. Mozilla Firefox

After searching the images acquired for Facebook, it was found that the user name and password for log in were present in images: Img4 through Img9. The username and password were preceded by the words 'email' and 'pass' which can be used as keywords for search.

As shown in Figure 3, the value set for user name is 'ipsita.chinky@gmail.com' and for password is 'who678%2C%3B'. The actual password entered was 'who678,;'. It could be concluded that the special symbols were converted into corresponding ASCII hex values, resulting in ',' as '%2C' and ';' as '%3B'. Hence while searching for passwords; care should be taken for the ASCII values stored. If the password contains letters from A-Z, a-z and numbers, no special characters, it could be easily identified.

It was observed that the username and password were available after logging out from facebook (Img7.img) and also after closing the browser (Img8 and Img9). However, the username and password was not found in the further images (Img10-Img13). This can be attributed to the fact that, after logging out from Sonicwall, the internet access permission got aborted and the Firefox window used for showing the user status information for Sonicwall closed. So, the process Firefox terminated completely resulting in the absence of the relevant data in images Img10-Img13.

Other useful information (except user name and password) like profile name, updates of the target user's friends etc. were also available for retrieval because the loaded pages were stored in RAM. The contents including user name and passwords were mostly in the memory space of firefox.exe and very few were in svchost.exe and kernel process.

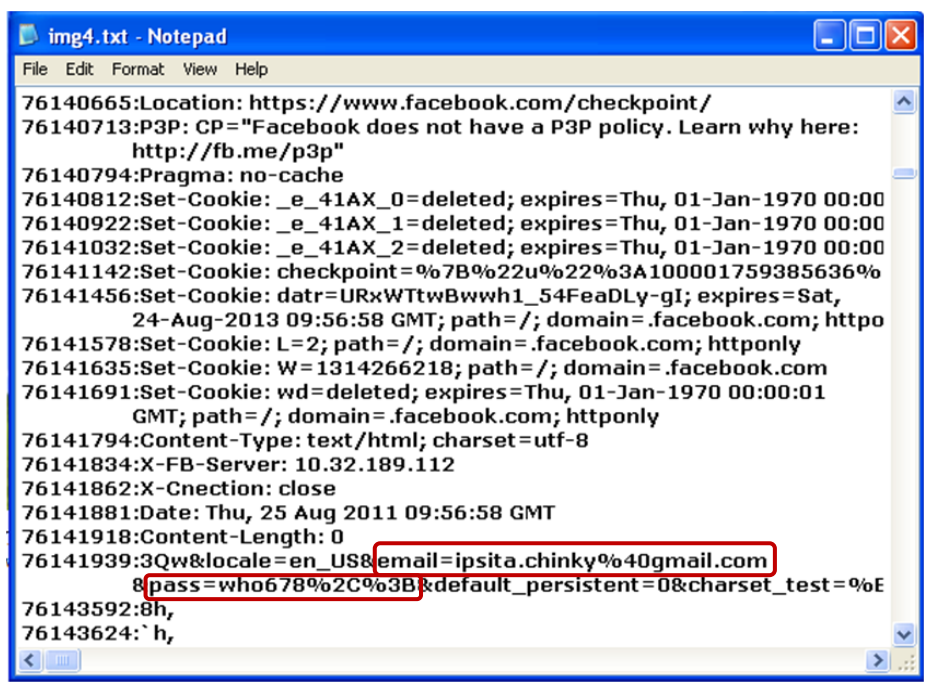

Figure 3. Snapshot of text file from Img4 taken for Facebook in Mozilla Firefox

### 4.2.2. Google Chrome

The user name and password for log in were present in images: Img4 through Img6. This was unlike Mozilla Firefox where the details were found in images: Img4 through Img9. However, the keywords for username and password were same as Firefox: 'email' and 'pass'.

As shown in Figure 4, the value set for user name is 'ipsita.chinky@gmail.com' and for password is 'berham!19'. The actual password entered was also 'berham!19'. So, unlike Mozilla Firefox, the special symbols were not converted into corresponding ASCII hex values.

Other useful information (except user name and password) like profile name and profile id of the user, updates of the target user's friends shown in news feed, friend requests, messages received, notifications, contacts as shown in the friends list, etc were also available for retrieval because the loaded pages were stored in RAM. The contents including user name and passwords were available till logging out from Facebook (Img6.img). No information was available in further images (Img7-Img13). These contents were in the memory space of chrome.exe and kernel process. It was concluded that even though the chrome.exe process was still running, logging out of Facebook prohibited access to these contents and only the user-id was available.

This brought out differences in information retrieval due to use of two different browsers. Each browser has separate settings and capabilities and hence, the information retrievable will vary from browser to browser for the same application.

```
img4.txt - Notepad
File Edit Format View Help
327658412:8@@
327658416:@@@@@@
327658660::://www.facebook.com/2
327658684:https://www.facebook.com/login.php?login_attempt=1
327658740:https://www.facebook.com/checkpoint/
327658792:http://www.facebook.com/
327658820:http://www.facebook.com/"
327658848:https://www.facebook.com/login.php
327658892:email
327658908:ipsita.chinky@gmail.com
327658960:pass
327658972:berham!19
327659016:text/html
327659032:69.171.229.74
327659128:https://www.facebook.com/checkpoint/d
327659204:https://www.facebook.com/login.php?login_attempt=1
327659308:_e_1MWL
327659368:http://www.facebook.com/
327659464:http://www.facebook.com/
327662592:DefaultConnectionSettings
327663276:X2@
327663746:A>a
327664030:B>}!
```

Figure 4. Snapshot of text file from Img4 taken for Facebook in Google Chrome

### 4.3. Gmail

#### 4.3.1. Mozilla Firefox

The username and password were retrievable from Gmail in a similar fashion to facebook. The details were available in images Img4 through Img9. The username and password were preceded by the words 'GAUSR=mail' and 'Passwd' which can be used as keywords for search.

A string, 'abc*%21123', very much similar to the entered password, 'abc*!123', was obtained for Gmail (Figure 5). It was found in images Img4 through Img9. In the password string, the special character '!' was converted into its ASCII hex value '21'. Thus, it was observed that if there were two hexadecimal digits after '%', the hexadecimal number should be converted to the associated special character.

In addition to username and password, other information like inbox contents and contacts were found. After logging into the account, the first page loaded contains inbox contents and some contacts available for chat. This ensured that the most recent inbox content and frequently used contacts could be retrievable. Similar to Facebook, the contents for Gmail were found in RAM while being logged into Sonicwall and stored in the memory space of firefox.exe, svchost.exe and kernel process.

Figure 5. Snapshot of text file from Img4 taken for Gmail in Mozilla Firefox

### 4.3.2. Google Chrome

Even for Google Chrome, the username and password for Gmail were retrievable. The details were available in images Img4 through Img6, but not after that i.e. in the images Img7- Img13. The username and password were preceded by the words 'Email' and 'Passwd' which can be used as keywords for search.

The password obtained for Gmail as in Figure 6, 'awesome^&28', was same as the value entered in the log in page. Thus, it was observed that there was no conversion of special characters into corresponding ASCII hex values.

In addition to username and password, other information like inbox contents and contacts were found. After logging into the account, the first page loaded contained inbox contents and some contacts available for chat. This ensured that the most recent inbox content and frequently used contacts could be retrievable. Similar to Facebook under Google chrome, the contents for Gmail were found in RAM while being logged into Gmail and were stored in the memory space of chrome.exe and kernel process.

This case was similar to that of Facebook, wherein, the data retrievable differed for the two browsers. This further strengthens the assumption that each browser is different and exhibits different levels of security for the same applications.

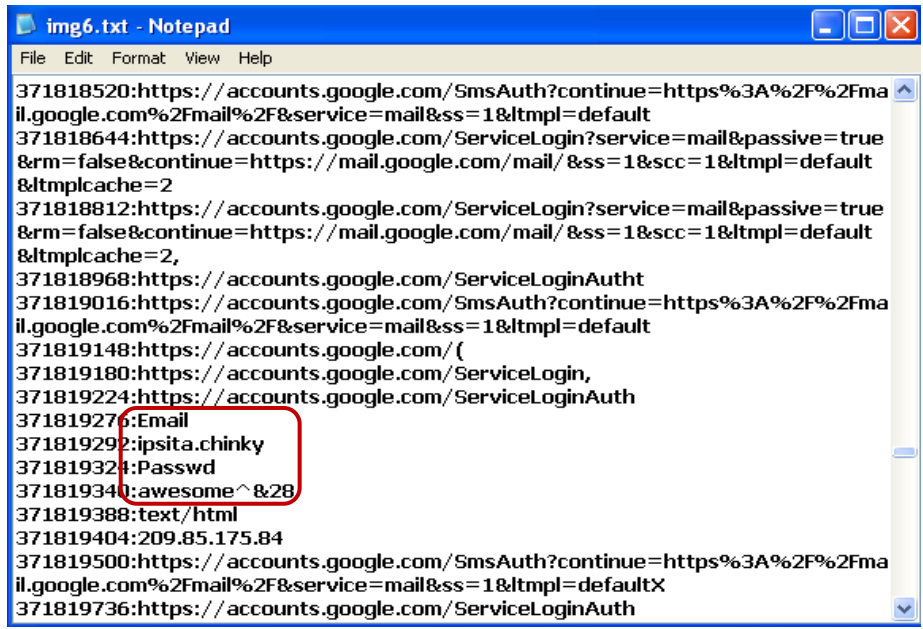

Figure 6. Snapshot of text file from Img6 taken for Gmail

## 4.4. IRCTC

### 4.4.1. Mozilla Firefox

For the application IRCTC, the user name and password were readily available and were found in images: Img4 through Img9. A snapshot highlighting the same is shown in Figure 7. The username and password were preceded by the words 'userName' and 'password' which can be used as keywords for easy search. Moreover, since there were no special characters used in the password, the password was available exactly without any encoding. The instances were in the memory space of firefox.exe and very few in that of svchost.exe and kernel process.

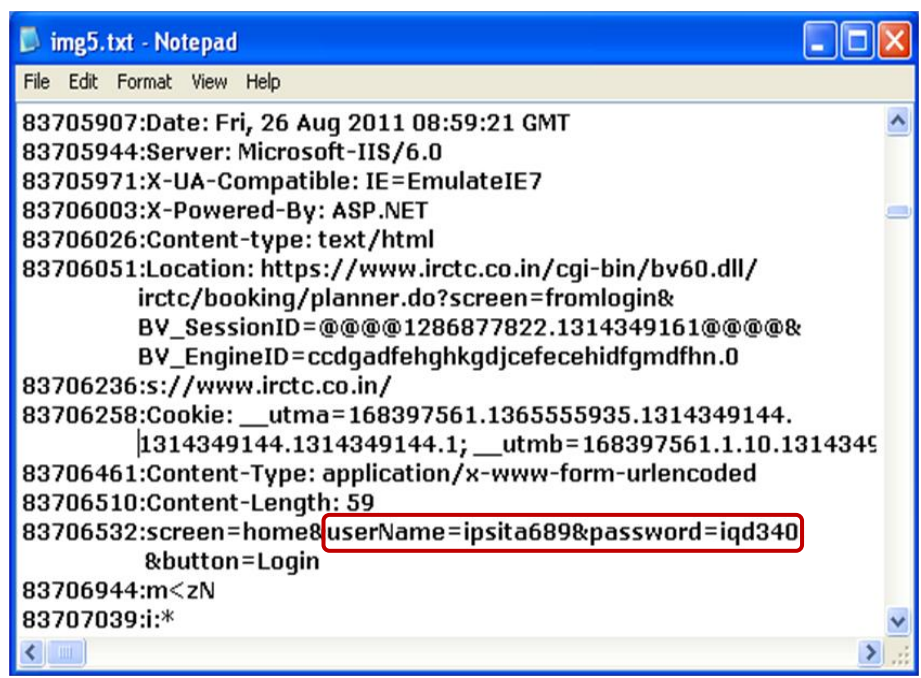

Figure 7. Snapshot of text file from Img5 taken for IRCTC in Mozilla Firefox

As explained for Gmail and Facebook, there were no presence of user name and password in images: Img10 through Img13 i.e., after logging out from Sonicwall.

**4.4.2. Google Chrome**

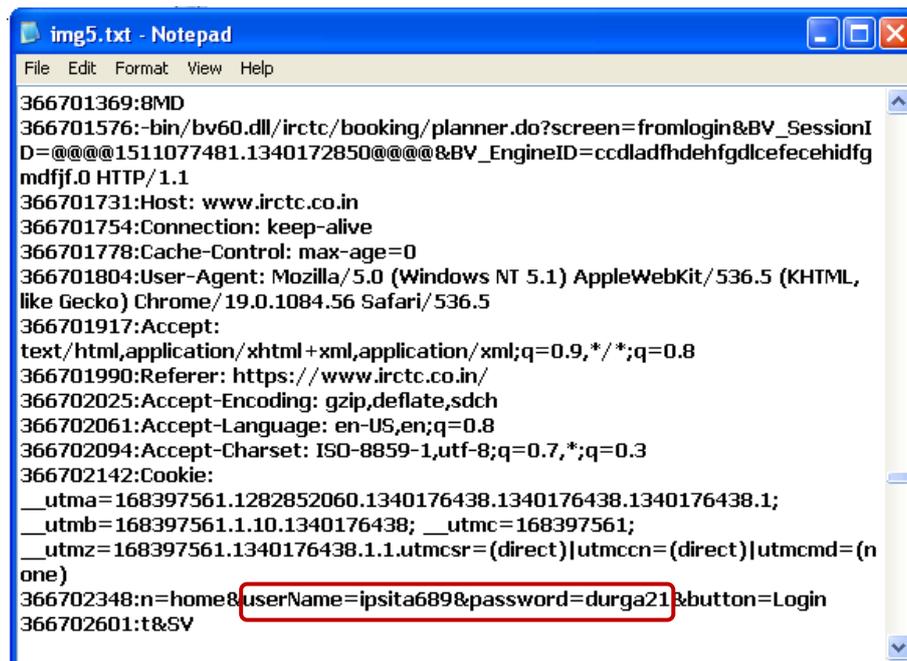

Figure 8. Snapshot of text file from Img5 taken for IRCTC in Google Chrome

The user name and password were readily available as preceded by the keywords 'userName' and 'password' and were found in images: Img4 through Img9. A snapshot highlighting user name and password is shown in Figure 8. The instances were in the memory space of chrome.exe and kernel process. Unlike Gmail and Facebook under Google Chrome, user name and password were present in the memory images taken after logging out from IRCTC website i.e. Img7-Img9. However, similar to Gmail and Facebook, the details were not found in Img10 through Img13 i.e., after logging out from Sonicwall. So the contents are recoverable until the process chrome.exe gets closed.

Unlike, Gmail and Facebook, this case gave similar results for both the browsers just like it did for the case of Sonicwall.

**4.5. SBI**

**4.5.1. Mozilla Firefox**

The same experimentation procedure was carried out for the internet banking site of State Bank of India. The password for logging into the application was not found in any of the memory images taken. The user name was seen as an isolated string, with no elements relating it to the log in page present nearby. On the contrary, the user name and password for Gmail, Facebook and IRCTC were near the website address. With no string related to the application page loaded and no preceding keywords to help in identification of the user name, it was difficult to identify the string as user name. However, the isolated string was in the memory space of Firefox.exe process, thereby opening up a possibility of association. But this is possible only when one application is opened. It will be a difficult task to figure this out if more than one application were used by the target user.

Apart from the user name, other useful information available was: account number, name of the account holder, bank branch code, name and branch of the bank. These contents were present only in images: Img4 through Img9. In the memory images taken after that, only the name of the website 'www.onlinesbi.com' was found. This shows that the online banking site is much better protected from acquisition as compared to the other websites.

### 4.5.2. Google Chrome

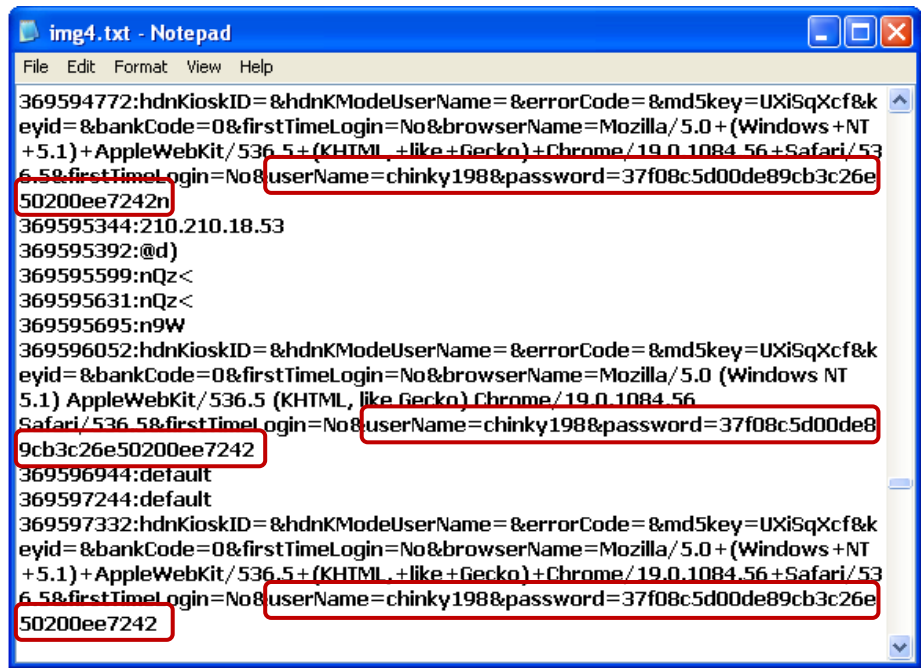

Figure 9. Snapshot of text file from Img4 taken for SBI in Google Chrome

In contrast to the results obtained for State Bank of India website for Mozilla Firefox, in Google Chrome, contents of the login page such as user id and password (in encrypted form) were available in memory images Img4.img through Img7.img i.e. after logging into the website till closing the browser. The keywords preceding user id and password were 'userName' and 'password' respectively (Figure 9). The actual password entered was 'pswd$%03', but the value stored in the variable password was '37f08c5d00de89cb3c26e50200ee7242' which was in encrypted form. In the locality of these contents, the url 'www.onlinesbi.com' was present, which enabled us to associate these data with the online banking website of SBI.

Apart from the user name, other useful information available was: account number, name of the account holder, bank branch code, name and branch of the bank. These contents were present in images: Img4 through Img7, in the memory space of chrome.exe and kernel process. In the memory images taken after that, only the name of the website 'www.onlinesbi.com' was found.

## 5. CONCLUSION AND FUTURE WORK

The approach followed in this paper is relevant to the existing global scenario where acquiring digital evidence holds primary importance in any investigation. Every browsing session of the target user leaves an imprint in the system memory and this has been exploited in this approach. It was possible to extract useful information from the memory images taken after the use of the application (without internet being severed). The application names or the words (different for different applications) preceding username and passwords, can be used as keywords for search. The information found out during analysis were username, password, list of contacts, mails, bank account number, name of the account holder etc. However, it was observed that the information was not available in the memory images taken after logging out of either the application for Gmail and Facebook (in Google Chrome) or the firewall (all other cases).

This represents the case of Live Acquisition wherein plenty of information can be retrieved about the state of the system in the recent past. Despite some limitations of live acquisition, it is impossible to ignore the importance of the contents of RAM. The utility of the approach is definitely on the higher side and is likely to find applications in a number of cases.

In this paper, two browsers are taken into consideration for conducting the experiments. The results obtained were different for different browsers in the case of Gmail and Facebook. The experiment can be extended to include more number of browsers to provide a more comprehensive conclusion to the results. Moreover, browsers are getting upgraded regularly and each version would have its own specific settings and features. The experiment can further include tests across various versions of every browser. This would ensure that the results are consistent across a number of versions of a number of browsers. Moreover, it would provide deep insights into the probability of retrieval of evidence irrespective of the browser and version the target user is using.

**Authors**

**R. Leela Velusamy** obtained her Bachelor degree in Electronics and Communication Engineering and Postgraduate degree in Computer Science and Engineering from Regional Engineering College (REC), Tiruchirappalli. She was awarded Ph.D. by the National Institute of Technology (NIT), Tiruchirappalli. Since 1989, she is in the teaching profession and currently she is an Associate professor in the Department of Computer Science and Engineering, National Institute of Technology, Tiruchirappalli, Tamil Nadu, India. Her research interests include Digital Forensics, Cryptography, Network Security and QoS routing.

**Ipsita Mohanty** obtained her B.Tech degree in Computer Science and Engineering from Silicon Institute of Technology, Bhubaneswar, Odisha, India. Currently, she is an MS (Research) scholar in the Department of Computer Science and Engineering, National Institute of Technology, Tiruchirappalli, Tamil Nadu, India. Her research interests include Digital Forensics and Network Security.